\documentclass[twocolumn]{article}      
\usepackage{authblk}
\usepackage{graphicx}
\usepackage[font={small}, labelfont=bf]{caption}
\usepackage[colorlinks=true,
            linkcolor=blue,
            citecolor=blue,
            urlcolor=blue]{hyperref}

\usepackage{program}
\usepackage{sectsty}
\sectionfont{\fontsize{12}{15}\selectfont}
\usepackage{geometry}
\title{STUDY OF THE MIGRATION OF EARTH-LIKE PLANETS
IN PLANETESIMAL DISKS AND THE FORMATION OF DEBRIS DISKS}  
\author{O. S. Oleynik*, V. V. Emel’yanenko  \vspace{-5pt}}      
\affil{Institute of Astronomy of the RAS, Moscow, Russia \vspace{-10pt}}
\affil{*e-mail: olgaoleynik93@gmail.com  \vspace{-10pt}}
\date{2025  \vspace{-10pt}}      

\begin{document}             

\twocolumn[
  \begin{@twocolumnfalse}
    \maketitle
\begin{abstract} The aim of this study is to investigate the interaction of Earth-mass planets with a planetesimal disk. It is shown that an Earth-mass planet, initially located near the inner boundary of the planetesimal disk, migrates into the disk. The depth of penetration of the planet into the disk is a random quantity determined by the angular momentum distribution of planetesimals approaching the planet. However, at a certain stage, the direction of the planet’s migration changes, and the planet returns to the inner boundary of the disk. During such reversible migration, the planet perturbs the orbits of planetesimals and increases their relative velocities in the region of the disk traversed during its migration. The relative velocities of planetesimals increase to values sufficient for their fragmentation in collisions. Our estimates show that, after the passage of an Earth-mass planet through the outer planetesimal disk, the mean relative velocities in the main part of the disk increase to values sufficient to disrupt monolithic basaltic planetesimals with sizes of ~40 km. Thus, the interaction of even a relatively low-mass planet (of order an Earth mass) with a planetesimal disk can lead to the formation of dust particles observed in outer debris disks.

 \vspace{1 cm}
\end{abstract}

  \end{@twocolumnfalse}
]
\section{Introduction}
\label{sec:1} 

The detection of numerous disks in the far-infrared wavelength range, whose luminosity is associated with the emission of dust heated by the star, appears to provide a direct opportunity to study dynamical processes occurring in planetesimal disks. For example, catalogue\footnote{https://www.astro.unijena.de/index.php/theory/catalog-of- resolved-debris-disks.html} contains data on 175 well-resolved debris disks, while more than a thousand candidate disks are reported in \cite{Cao2023}. Many such structures are observed around main-sequence stars older than 10 Myr \cite{Pearce2022,Boccaletti2023}. Therefore, it can be assumed that these disks contain only a small fraction of the primordial protoplanetary material or do not contain it at all.

The existence of debris disks is generally attributed to planetary perturbations of the relative velocities of planetesimals, leading to destructive collisions and the production of dust particles (see, e.g., \cite{Mustill2009}). However, the dust component of a debris disk must be continuously replenished, since the lifetime of the observed dust is shorter than the age of the star–disk system \cite{Backman1993,MoroMartin2007}. The lifetime of dust particles is determined by a combination of several processes. Dust is removed from the vicinity of the star by radiation pressure, spirals inward towards the star due to the Poynting–Robertson effect, and is destroyed in collisions. The dominant mechanism of dust replenishment is most likely collisions between large planetesimals \cite{Krivov2021,Lovell2022,Lestrade2025}.

For dust production via collisions, sufficiently high relative velocities of planetesimals are required \cite{Krivov2010}. At present, the role of planets in the formation of debris disks, as well as in shaping various observed features of dust distributions within disks, remains an active topic and is widely discussed in the literature \cite{Pearce2022,Boccaletti2023,Gaspar2023}.

As in the case of the Kuiper belt in the Solar System, a typical debris disk most likely represents a ring of dust-producing planetesimals located at distances of several tens of au from the star (see, e.g., \cite{Boccaletti2023,Akeson2009,Wyatt2018,Mesa2021}). Such disks are referred to as “outer” disks \cite{Krivov2010}. The present study is aimed at investigating the processes leading to the formation of an outer debris disk around a solar-type star. Objects of this kind are fairly common. Approximately 28 \% of nearby F, G, and K-type stars are surrounded by debris disks \cite{Sibthorpe2018}.

The key issue in dust production is the mechanism responsible for perturbing the orbits of planetesimals. Two mechanisms for stirring planetesimal disks are most commonly considered:

1) the process of “self-stirring”, in which large planetesimals excite smaller ones \cite{Kenyon2001,Kenyon2008,Kenyon2010,Kennedy2010,Krivov2018};

2) the action of secular perturbations from a giant planet \cite{Mustill2009}.

However, as emphasised in \cite{Costa2024}, these mechanisms are not universal. The first mechanism does not require the presence of planets in the disk but implies unrealistically high disk masses \cite{Pearce2022,Krivov2021,Najita2022}. The second mechanism operates only in the case of planets on highly eccentric orbits \cite{Mustill2009}. In addition, the self-gravity of the disk significantly reduces the efficiency of orbital excitation of disk particles by secular perturbations \cite{Sefilian2024}.

We consider a new mechanism for stirring a planetesimal disk, in which the orbits of planetesimals are perturbed by a planet of relatively low mass (of order an Earth mass), initially located near the inner boundary of the disk. Although Earth-like planets are absent among the known planets in the outer Solar System and have not been detected at large distances from stars in exoplanetary systems, there are strong reasons to assume their existence at early stages of planetary system formation. Numerical studies of planetary system formation indicate that, together with giant planets, bodies with masses ranging from fractions of an Earth mass to several Earth masses are always formed (see, e.g., \cite{Safronov1966,Wetherill1992,Petit1999,Fernandez1996,Gladman2002}). It is also noteworthy that super-Earths constitute a significant fraction of detected exoplanets, and their occurrence rate increases with decreasing mass (although current observational methods allow this trend to be established only for objects with orbital periods shorter than 100 days) \cite{Ananyeva2022}.

\section{Model}
\label{sec:2} 

We study a dynamical process at an early stage of planetary system evolution after the dispersal of the gas disk. The main objective of this work is to identify the dynamical features of the interaction between a migrating planet and a planetesimal disk. Therefore, we consider a model in which, at the initial moment, the planet is located on a nearly circular orbit close to the inner edge of the disk (at a distance of one Hill radius). This approach allows us to reduce computational costs while focusing on the primary objective of the study, without modelling the prior evolutionary history of the planet.

We modelled the motion of a planet with different masses: $0.5\,M_\oplus$, $1\,M_\oplus$, and $5\,M_\oplus$, where $M_\oplus$ is the Earth’s mass. The initial eccentricity of the planet was set to $0.01$, and its motion initially lay in the reference plane. At the initial moment, planetesimals also move on nearly circular orbits. The initial eccentricities $e$ and inclinations $i$ of planetesimal orbits were uniformly distributed in the intervals $(0,\,0.01)$ and $(0^\circ,\,0.5^\circ)$, respectively. The semimajor axes of planetesimals were distributed according to a power law $a^{-1}$ in the range from $30$ to $40\,\mathrm{au}$, where $a$ is the semimajor axis. Such initial conditions are natural for a system formed in a gas--dust disk, and similar assumptions have been adopted in many studies of planet migration in planetesimal disks (see, e.g., \cite{Morbidelli2009,Nesvorny2012,Kaib2016,Griveaud2024}).

In this work, we considered models with planetesimal disk masses of $20\,M_\oplus$ and $40\,M_\oplus$, consistent with estimates of the disk mass beyond Neptune’s orbit in the Nice model (see, e.g., \cite{Nesvorny2012,Griveaud2024,Crida2009}). In the numerical experiments, the planetesimal disk contained $8982$ equal-mass bodies. This number of objects is determined by the numerical integration method, similar to that used in \cite{Gomes2004}. The integration time of the equations of motion was $10$--$15\,\mathrm{Myr}$. The equations of motion were solved numerically in the framework of the $N$-body problem using a symplectic integrator \cite{Emelyanenko2007}. We adopted a widely used approach \cite{Nesvorny2012,Griveaud2024}, in which the gravitational interaction between the planet and planetesimals is fully taken into account, while planetesimals do not interact gravitationally with each other.

\section{Planet migration}

\begin{figure}[h!]
  \vspace*{-0.2cm}
    \centering
    \hspace*{-15pt}\includegraphics[width=1.15\columnwidth]{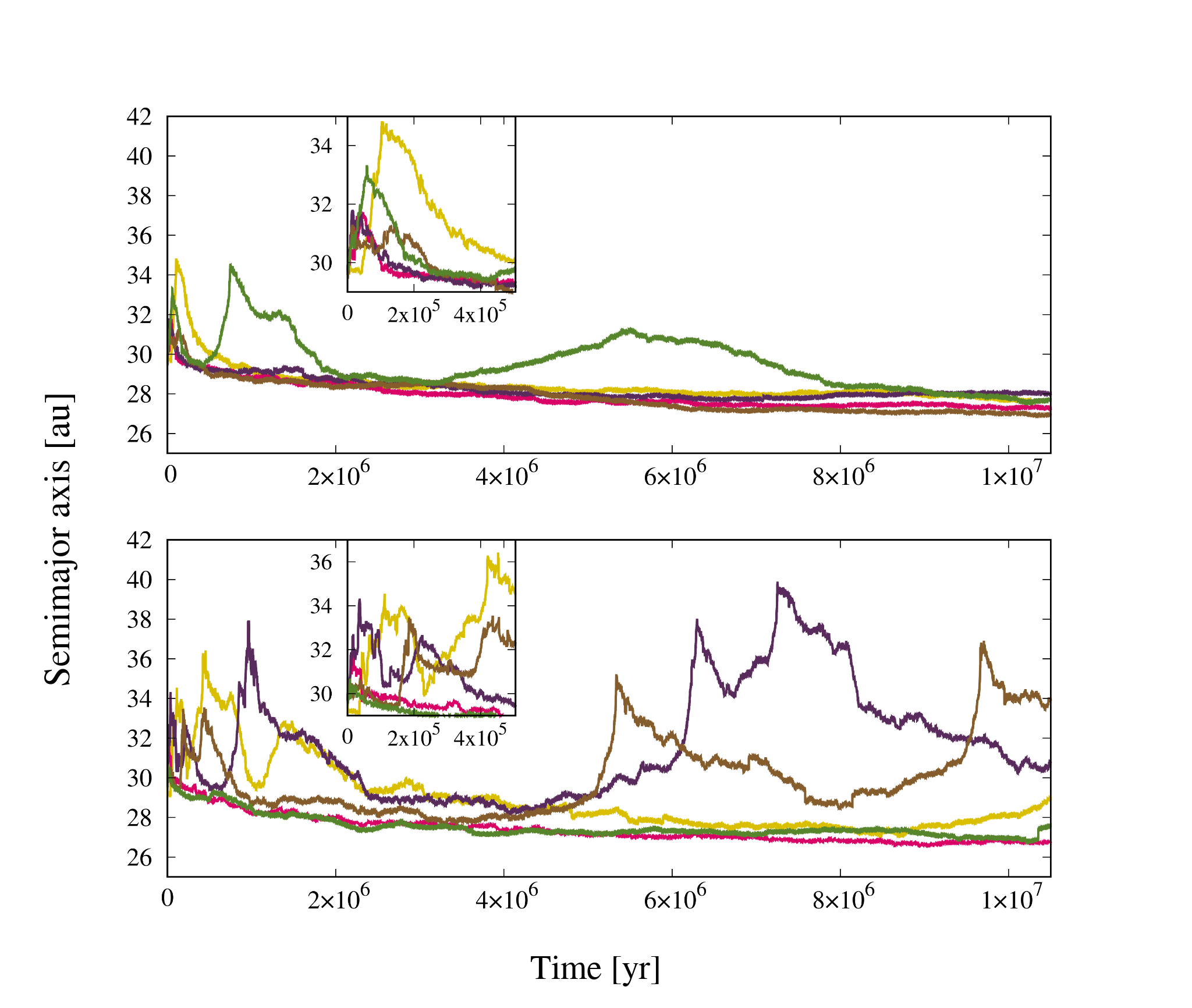}
    \vspace*{-0.7cm}
    \caption{Evolution of the semimajor axis of a $0.5\,M_\oplus$ planet in planetesimal disks with masses of $20\,M_\oplus$ (top panel) and $40\,M_\oplus$ (bottom panel). The same curves are also shown in a zoomed-in view over the time interval $0$–$0.5\,\mathrm{Myr}$.}
    \label{fig:fig1}
\end{figure}

\begin{figure}[h!]
\vspace*{-0.74cm}
    \centering
    \hspace*{-15pt}\includegraphics[width=1.15\columnwidth]{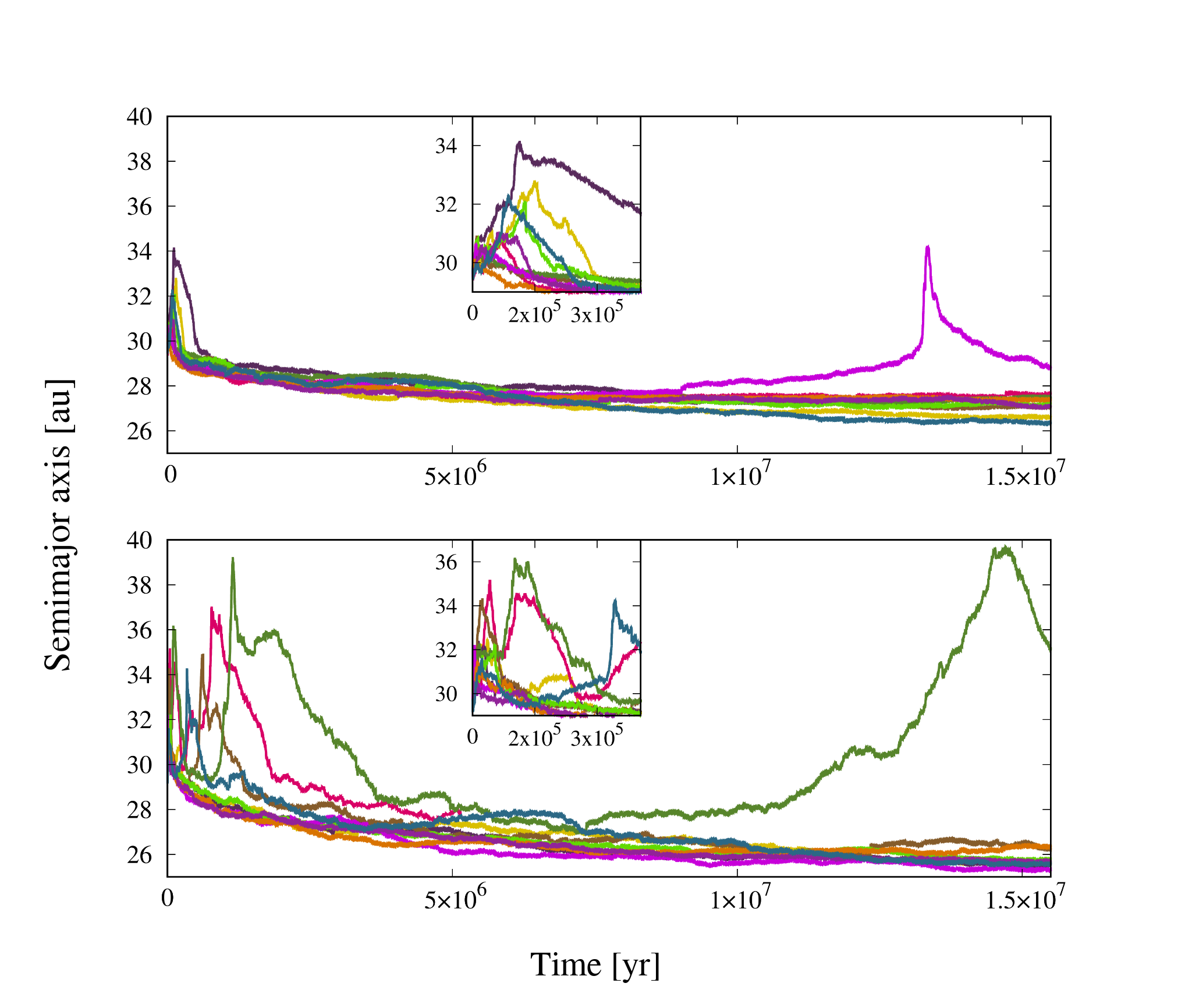}
    \vspace*{-0.7cm}
    \caption{Evolution of the semimajor axis of a $1\,M_\oplus$ planet in planetesimal disks with masses of $20\,M_\oplus$ (top panel) and $40\,M_\oplus$ (bottom panel) over $15\,\mathrm{Myr}$. The same curves are also shown in a zoomed-in view over the time interval $0$–$0.4\,\mathrm{Myr}$.
}
    \label{fig:fig2}
\end{figure}

\begin{figure}[h!]
 \vspace*{-0.4cm}
    \centering
    \hspace*{-15pt}\includegraphics[width=1.15\columnwidth]{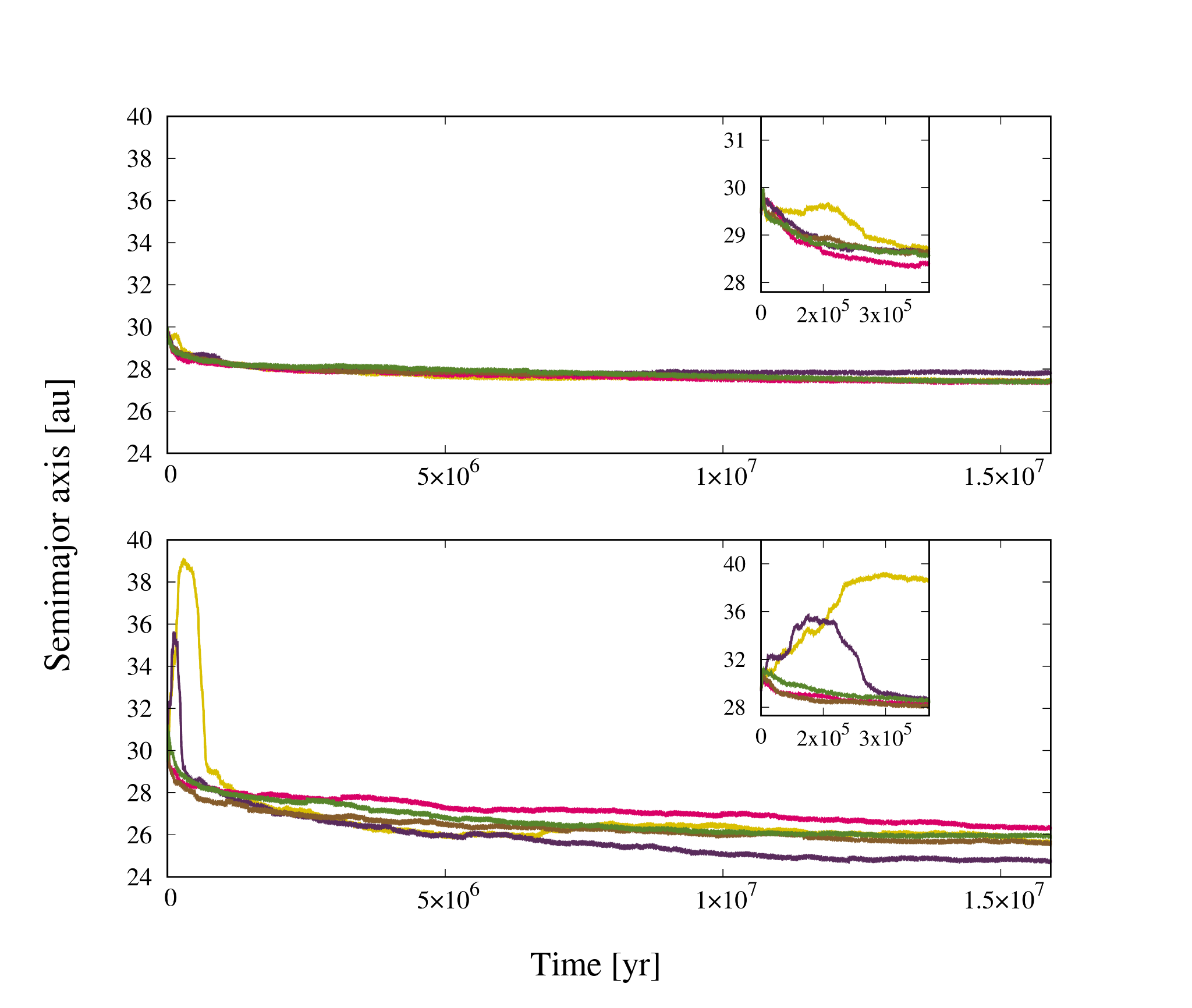}
    \vspace*{-0.7cm}
    \caption{Evolution of the semimajor axis of a $5\,M_\oplus$ planet in planetesimal disks with masses of $20\,M_\oplus$ (top panel) and $40\,M_\oplus$ (bottom panel) over $15\,\mathrm{Myr}$. The same curves are also shown in a zoomed-in view over the time interval $0$–$0.35 \,\mathrm{Myr}$.}
    \label{fig:fig3}
\end{figure}

Migration of the planet, i.e. a change in its semimajor axis, occurs as a result of gravitational interaction with the planetesimal disk. For each value of the planet mass, several runs were performed in which the initial orbits of planetesimals were different realisations of random distributions with identical parameters (see Section~\ref{sec:2}). Figures \ref{fig:fig1} -- \ref{fig:fig3} show the migration of planets in disks with masses of $20\,M_\oplus$ (top panel) and $40\,M_\oplus$ (bottom panel). The planetary motion is seen to be chaotic, leading to a wide range of orbital evolution outcomes.

The planet’s motion relative to the disk can be complex; however, initially it is always directed outward, toward the disk. This is because the planet starts near the inner edge of the disk in all simulations and therefore interacts only with planetesimals exterior to its orbit. Consequently, it scatters them inward, gains angular momentum, and migrates into the disk.

Although in all simulations the planet initially migrates into the disk, its subsequent evolution varies from run to run. In most cases, it reverses direction before reaching the middle of the disk and then migrates inward, toward the star, crossing the initial inner edge. It subsequently remains there, with its semimajor axis changing only slightly until the end of the integration.

There are cases in which, after the first entry into the disk, the planet repeatedly changes the direction of migration while remaining within the disk. The frequency of this behaviour increases with decreasing ratio of the planet mass to the disk mass. For a mass ratio of $0.0125$, this effect occurs in three out of five runs; for $0.025$, in two out of ten runs; and for $0.125$, it is not observed.

However, there are also cases in which the planet, after leaving the disk, re-enters it. For a fixed disk mass, the frequency of this behaviour decreases with increasing planet mass, since a more massive planet efficiently clears the vicinity of its orbit when located near the edge of the disk.

Particularly interesting are cases in which the planet migrates through the entire disk. In such cases, the direction of migration always reverses. This effect, occurring near the outer edge of the disk, has been reported previously in the literature \cite{Gomes2004,Emelyanenko2010}. It should be noted that, in the numerical experiments, such passages occurred primarily in models with more massive disks and for more massive planets.

\section{Evolution of planetesimal disk}

Migrating planets interact with disk particles, modifying the distribution of their orbits. Thus, the motion of a planet within the disk determines its structure. As a result of the diversity of planetary migration in planetesimal disks, a variety of structural features can arise.

\begin{figure}[h!]
 \vspace*{-0.4cm}
    \centering
    \hspace*{-10pt}\includegraphics[width=1.1\columnwidth]{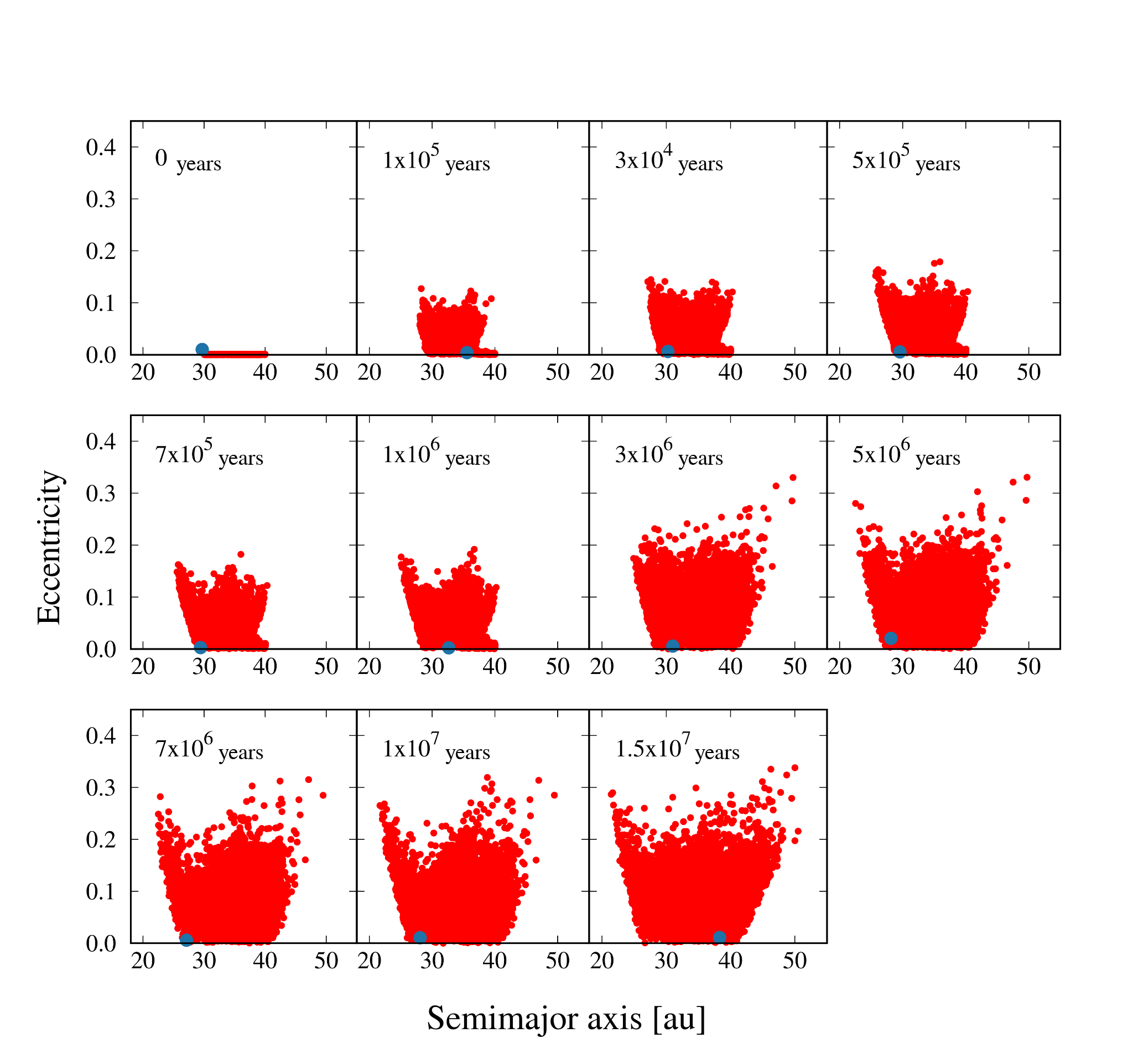}
    \vspace*{-0.7cm}
    \caption{Semimajor axes and eccentricities of planetesimals at different epochs during the migration of a $1\,M_\oplus$ planet in a $40\,M_\oplus$ disk. Red dots correspond to planetesimals, while the blue dot denotes the planet.}
    \label{fig:fig4}
\end{figure}

\begin{figure}[h!]
 \vspace*{-0.25cm}
    \centering
    \hspace*{-10pt}\includegraphics[width=1.1\columnwidth]{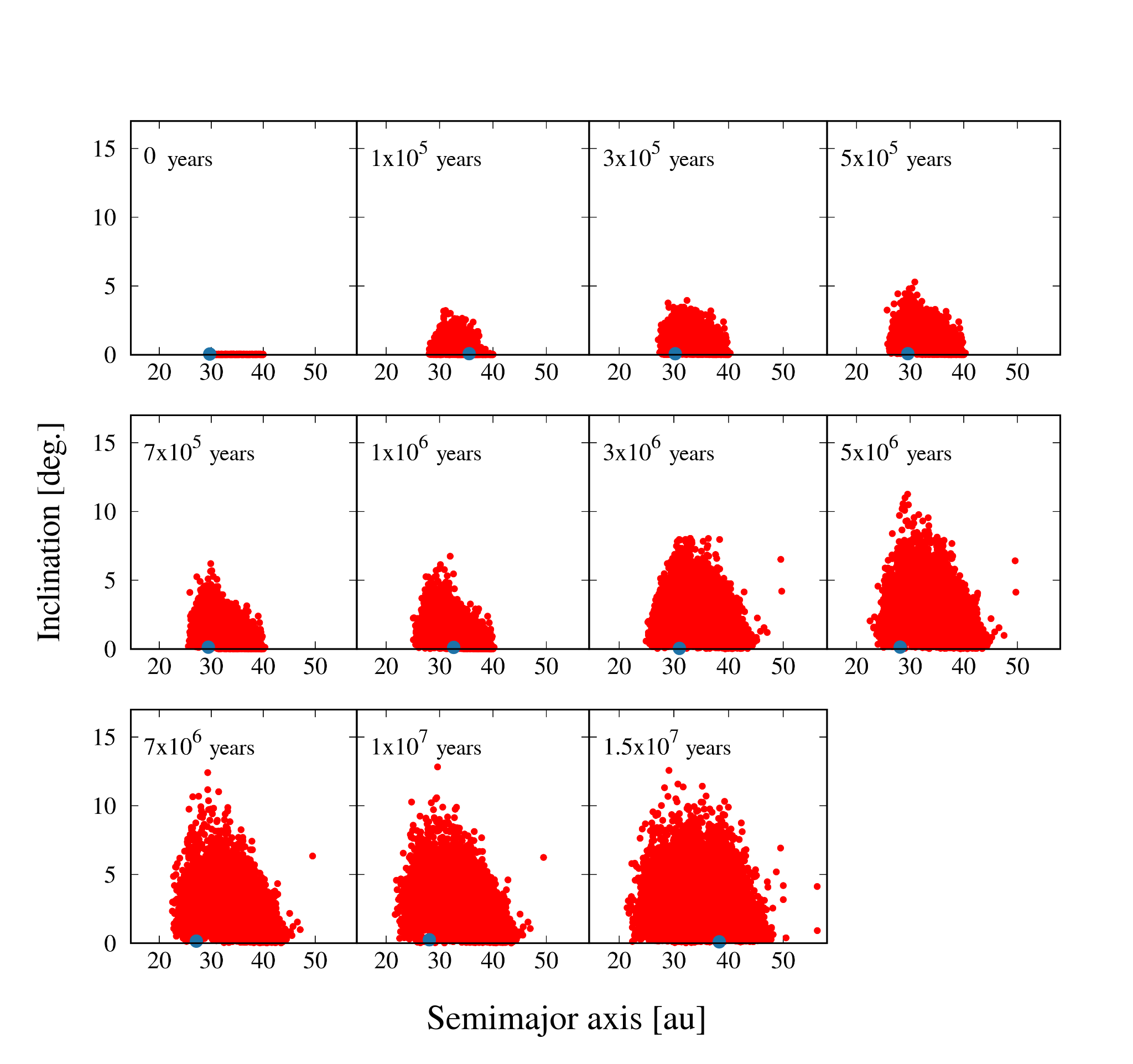}
    \vspace*{-0.7cm}
    \caption{Semimajor axes and inclinations of planetesimals at different epochs during the migration of a $1\,M_\oplus$ planet in a $40\,M_\oplus$ disk. Red dots correspond to planetesimals, while the blue dot denotes the planet.}
    \label{fig:fig5}
\end{figure}

\begin{figure}[h!]
 \vspace*{-0.7cm}
    \centering
    \hspace*{-10pt}\includegraphics[width=1.1\columnwidth]{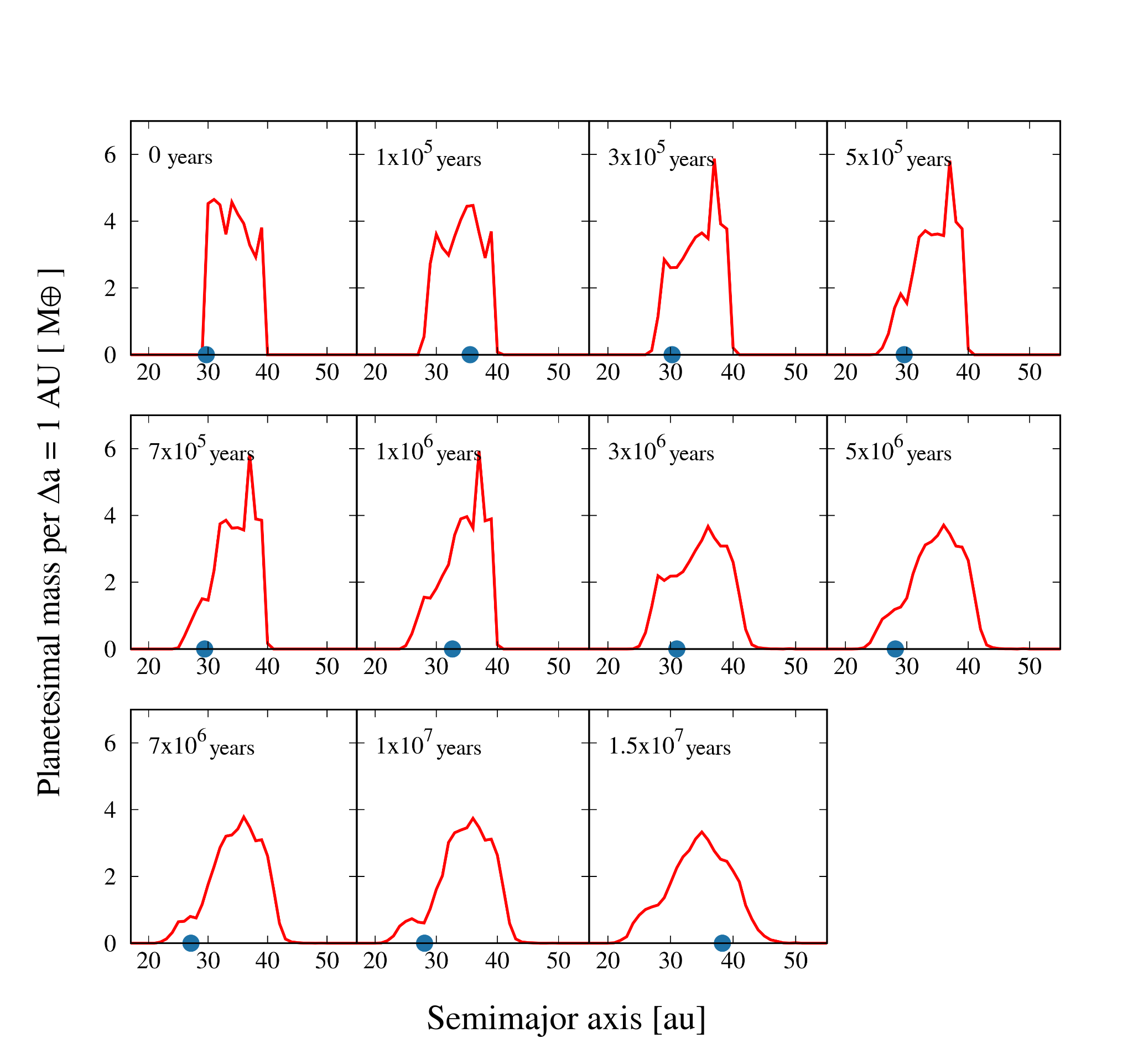}
    \vspace*{-0.7cm}
    \caption{Evolution of the radial mass distribution of planetesimals as a function of semimajor axis during the migration of a $1\,M_\oplus$ planet in a $40\,M_\oplus$ disk. The mass is computed in bins of width $1\,\mathrm{au}$. The blue dot indicates the semimajor axis of the planet.}
    \label{fig:fig6}
\end{figure}

We consider the evolution of the disk for one of the migration scenarios in which the planet passes through the entire disk (green curve in the lower panel of Fig.~\ref{fig:fig2}). In this case, a planet with a mass of $1\,M_\oplus$ evolves in a disk with a mass of $40\,M_\oplus$. As shown in Figs.~\ref{fig:fig4}--\ref{fig:fig6}, planetary perturbations lead to an expansion of the planetesimal disk boundaries.

Up to time $t = 3 \times 10^6$ yr, the eccentricities (Fig.~\ref{fig:fig4}) and inclinations (Fig.~\ref{fig:fig5}) of particles in the outer part of the disk remain relatively small. Indeed, as shown in the lower panel of Fig.~\ref{fig:fig2} (green curve), by this time the planet completes its first passage through the entire disk. As a result, by the end of the integration the orbits of the majority of planetesimals are significantly modified (the maximum eccentricities exceed $0.3$, and the inclinations exceed $10^\circ$).

Close encounters with the planet provide the dominant contribution to the orbital evolution of planetesimals. The first passage of the planet through the disk plays the key role in shaping the particle distribution, when planetesimals experience low-velocity encounters with the planet. As seen in Figs.~\ref{fig:fig4} and \ref{fig:fig5}, from $t = 5 \times 10^6$ yr onwards, the shapes of the eccentricity and inclination distributions do not change significantly.

As shown in Fig.~\ref{fig:fig4}, the passage of the planet through the disk alters the mass distribution. As a result, a significant fraction of the planetesimal mass moves beyond the initial disk boundaries at $30$ and $40\,\mathrm{au}$. Over the considered time interval ($15\,\mathrm{Myr}$), the planet spends more time near the inner edge of the disk than near the outer edge. As seen in the lower panel of Fig.~\ref{fig:fig2} (green curve), the planet remains close to the initial inner edge from $t = 3 \times 10^6$ to $t = 1 \times 10^7$ yr. Consequently, it scatters particles near the inner edge more efficiently than near the outer edge (Fig.~\ref{fig:fig6}).

After the second entry into the disk (after $t = 1 \times 10^7$ yr), as the planet migrates deeper into the disk and reverses near its outer edge, it again modifies the mass distribution near the outer edge. As seen in Fig.~\ref{fig:fig6}, at $t = 1.5 \times 10^7$ yr the mass at $a > 40\,\mathrm{au}$ increases; however, this change is relatively small. The mass distribution does not undergo significant changes after $t = 5 \times 10^6$ yr, and it can be concluded that it is primarily established during the first passage of the planet through the disk.

As seen in Fig.~\ref{fig:fig6}, by the end of the integration most of the mass remains within the initial disk boundaries ($30$--$40\,\mathrm{au}$).

We now consider another migration scenario in which the planet passes through the entire disk (yellow curve in the lower panel of Fig.~\ref{fig:fig3}). In this case, a planet with a mass of $5\,M_\oplus$ migrates in a disk with a mass of $40\,M_\oplus$. The planet passes through the disk only once, after which it remains near the inner edge until the end of the integration.

\begin{figure}[h!]
 \vspace*{-0.4cm}
    \centering
    \hspace*{-10pt}\includegraphics[width=1.1\columnwidth]{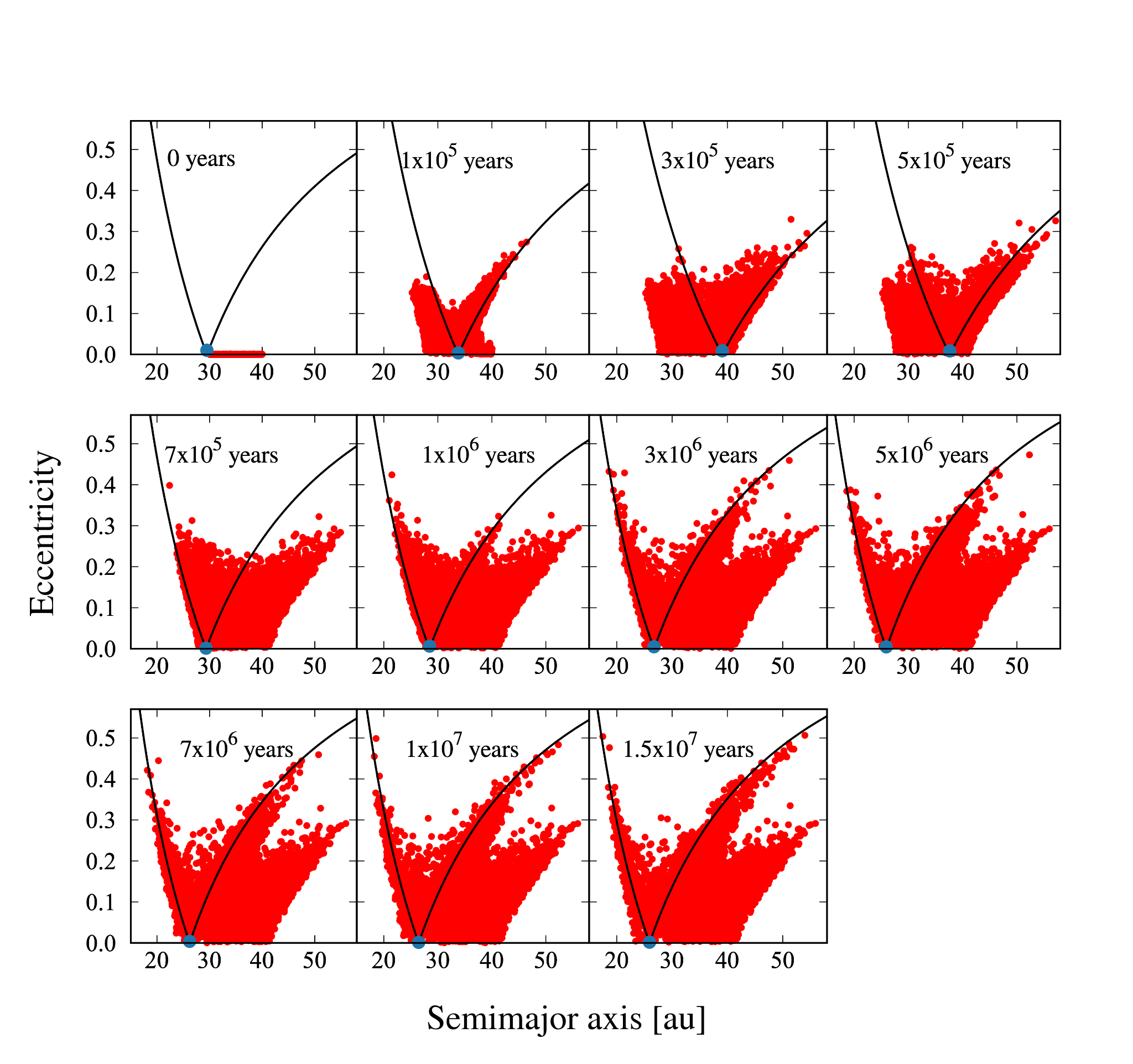}
    \vspace*{-0.7cm}
    \caption{Semimajor axes and inclinations of planetesimals at different epochs during the migration of a $5\,M_\oplus$ planet in a $40\,M_\oplus$ disk. Red dots correspond to planetesimals, while the blue dot denotes the planet.}
    \label{fig:fig7}
\end{figure}

The increase in eccentricity (Fig.~\ref{fig:fig7}) in this case is larger than in the case shown in Fig.~\ref{fig:fig4} (migration of a $1\,M_\oplus$ planet). As it migrates through the disk, the more massive planet perturbs particle orbits more efficiently. As a result, by the end of the integration the orbits of most planetesimals are significantly modified (the maximum eccentricities reach $\sim 0.5$).

In addition, a characteristic feature of the eccentricity distribution as a function of semimajor axis is clearly visible, namely the presence of “wings” --- lines along which particles are aligned in Fig.~\ref{fig:fig7}. This feature arises because the strongest orbital changes occur during low-velocity encounters between planetesimals and the planet near pericentre or apocentre. In such encounters, the eccentricity increases, while the pericentre or apocentre remains close to the value of the planet’s semimajor axis at the moment of encounter (corresponding to the right and left “wing”, respectively).

In the case of disk evolution driven by the migration of a $5\,M_\oplus$ planet (Fig.~\ref{fig:fig7}), the “wings” are more pronounced than for a $1\,M_\oplus$ planet (Fig.~\ref{fig:fig4}). This is due to two factors. First, the larger mass of the planet. Second, in this scenario the $5\,M_\oplus$ planet spends about $t = 3 \times 10^6$ yr near the outer edge of the disk and about $t = 1.5 \times 10^7$ yr near the inner edge (yellow curve in the lower panel of Fig.~\ref{fig:fig3}).

\section{Formation of debris disk}

A debris disk is observed through the emission of dust. The production of dust requires that the relative velocities of disk particles reach values sufficient for the disruption of planetesimals. During disruption, planetesimals produce smaller fragments, including observable dust, thereby initiating a collisional cascade \cite{WyattDent2002,Pearce2022}. The velocity required for the disruption of planetesimals (the fragmentation velocity) depends on the sizes of the colliding planetesimals and their composition \cite{Krivov2005,Schuppler2015}.

One of the most commonly used approaches for describing the stochastic process of object disruption in a planetesimal disk is based on the mean relative velocity \cite{Greenberg1978}. The problem of determining the mean relative velocity in a given planetesimal disk is non-trivial \cite{Lissauer1993}. In this work, we use a widely adopted expression to estimate the mean relative velocity in the disk \cite{Wyatt2007,Lissauer1993}:
\begin{equation}
v_{\mathrm{rel}} = v_{\mathrm{K}} \sqrt{\frac{5}{4}\langle e^2 \rangle + \langle i^2 \rangle},
\end{equation}
where $v_{\mathrm{K}}$ is the Keplerian velocity, and $\langle e^2 \rangle$ and $\langle i^2 \rangle$ are the mean squared eccentricity and inclination, respectively.

\begin{figure}[h!]
 \vspace*{-0.4cm}
    \centering
    \hspace*{-10pt}\includegraphics[width=1.1\columnwidth]{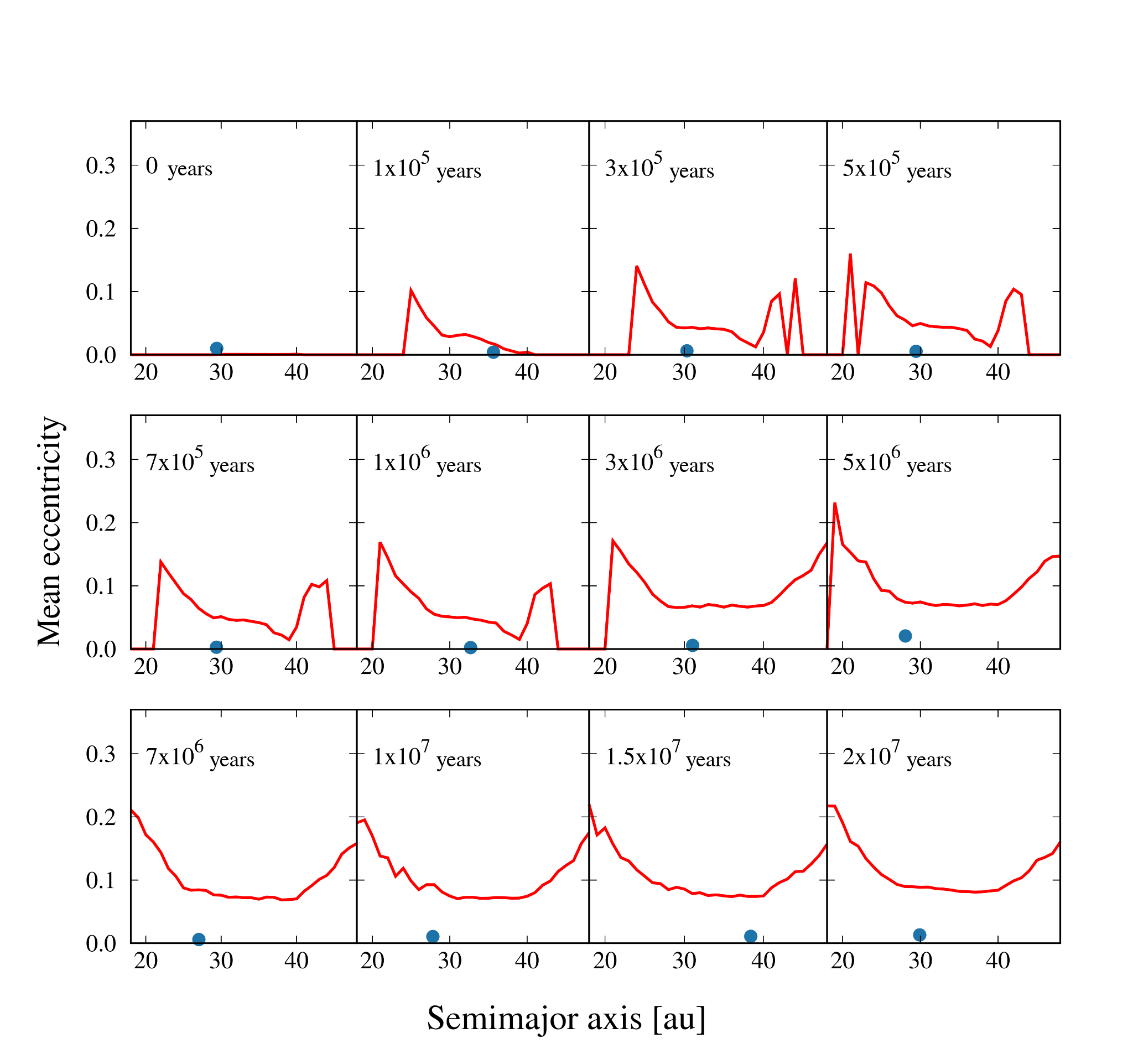}
    \vspace*{-0.7cm}
    \caption{Evolution of the radial distribution of the mean eccentricity of planetesimals during the migration of a $1\,M_\oplus$ planet in a $40\,M_\oplus$ disk. The blue dot denotes the planet.}
    \label{fig:fig8}
\end{figure}

\begin{figure}[t]
 \vspace*{-0.4cm}
    \centering
    \hspace*{-10pt}\includegraphics[width=1.1\columnwidth]{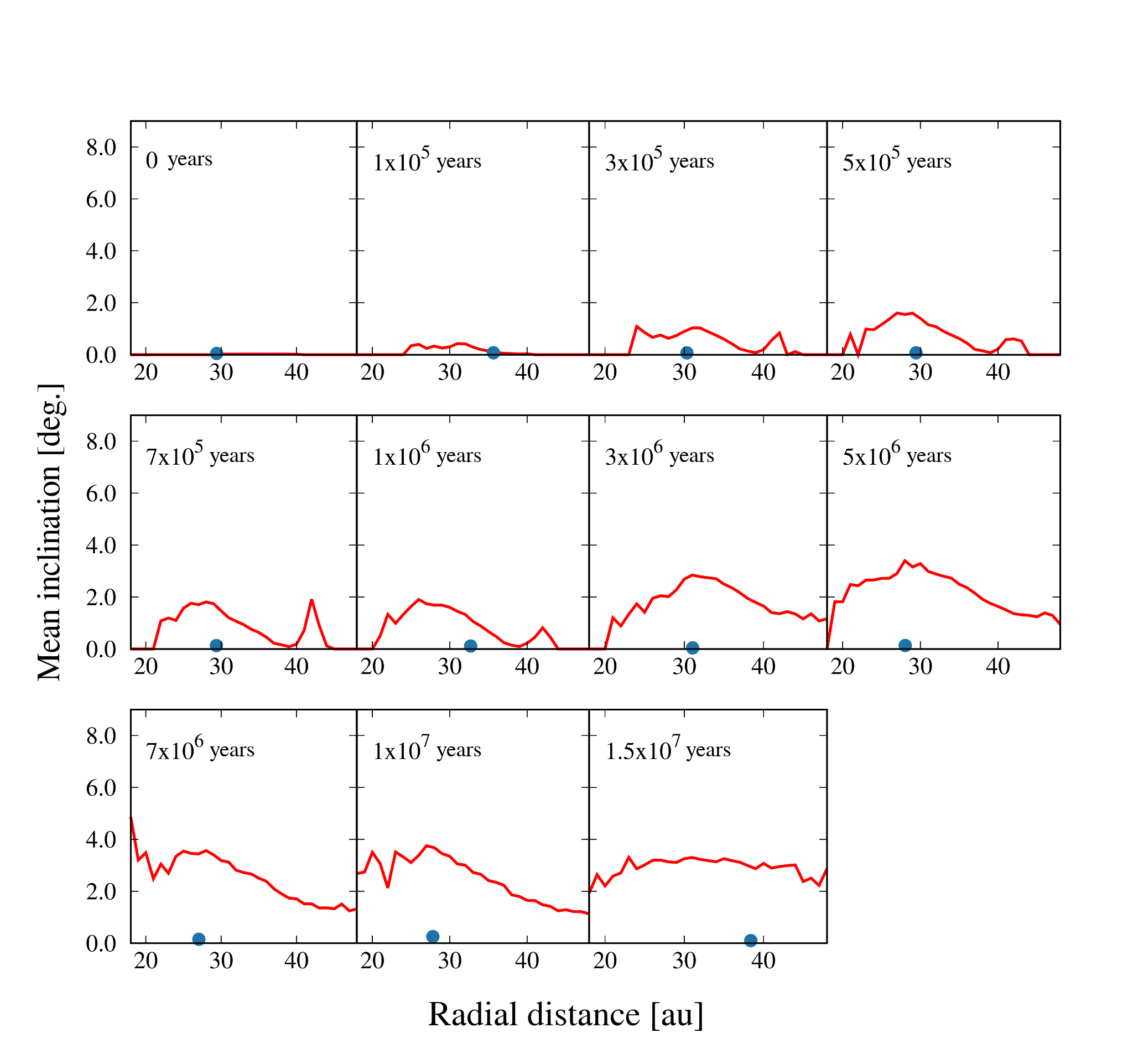}
    \vspace*{-0.7cm}
    \caption{Evolution of the distribution of the mean inclination of planetesimals during the migration of a $1\,M_\oplus$ planet in a $40\,M_\oplus$ disk. The blue dot denotes the planet.}
    \label{fig:fig9}
\end{figure}

\begin{figure}[h!]
 \vspace*{-0.2cm}
    \centering
    \hspace*{-10pt}\includegraphics[width=1.1\columnwidth]{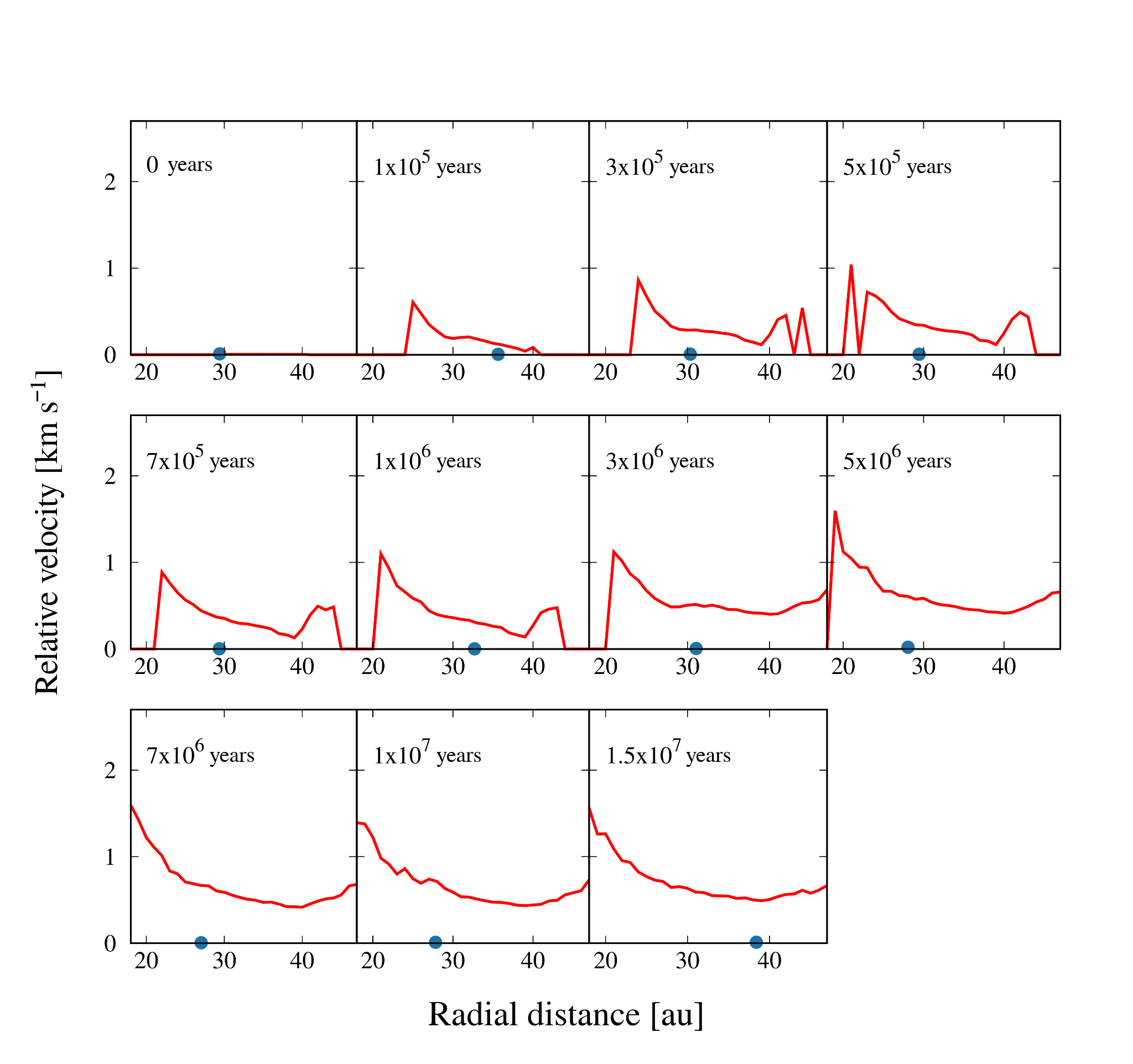}
    \vspace*{-0.7cm}
    \caption{Evolution of the relative velocity as a function of distance during the migration of a $1\,M_\oplus$ planet in a $40\,M_\oplus$ disk. The blue dot indicates the distance of the planet from the star.}
    \label{fig:fig10}
\end{figure}

Figures~\ref{fig:fig8} and \ref{fig:fig9} show the distributions of the mean eccentricities and inclinations of planetesimals in the case of migration of a $1\,M_\oplus$ planet in a disk with a mass of $40\,M_\oplus$ (green curve in the lower panel of Fig.~\ref{fig:fig2}). The eccentricities, inclinations, and relative velocity are evaluated within heliocentric distance intervals of $1\,\mathrm{au}$, while the Keplerian velocity is calculated at the centre of each interval. The evolution of the relative velocity of planetesimals in the disk is shown in Fig.~\ref{fig:fig10}.

In this work, the following expression is used to estimate the fragmentation velocity for collisions between two monolithic basaltic planetesimals of equal size \cite{Costa2024}:
\begin{equation}
v_{\mathrm{frag}} = 17.5 \left(\frac{D}{\mathrm{km}}\right)^{0.93}\ \mathrm{m\,s^{-1}},
\end{equation}
where $D$ is the size of large planetesimals ($D > 1\,\mathrm{km}$).

In the case of migration of a $1\,M_\oplus$ planet (green curve in the lower panel of Fig.~\ref{fig:fig2}), the mass distribution in the disk shown in Fig.~\ref{fig:fig12} does not change significantly from $t = 3 \times 10^6$ yr onwards. The same applies to the relative velocities (Fig.~\ref{fig:fig10}). We take this moment to mark the onset of the collisional cascade in the disk.

\begin{figure}[h!]
 \vspace*{-0.2cm}
    \centering
    \hspace*{-15pt}\includegraphics[width=1.0\columnwidth]{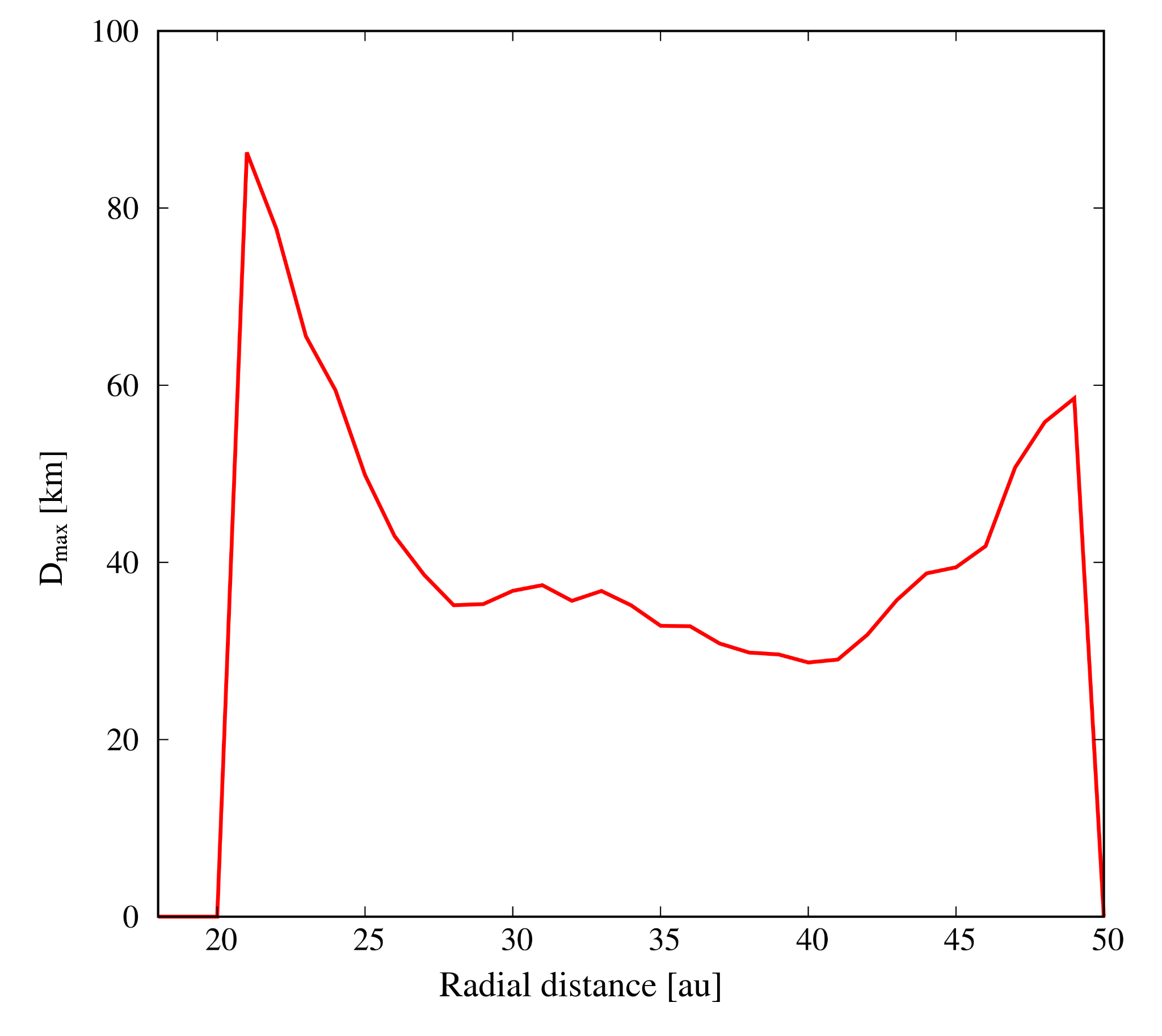}
    \vspace*{-0.2cm}
    \caption{Maximum planetesimal diameter $D_{\max}$ as a function of distance at $3\,\mathrm{Myr}$ after the start of the integration for a $1\,M_\oplus$ planet in a $40\,M_\oplus$ disk.}
    \label{fig:fig11}
\end{figure}

\begin{figure}[h!]
 \vspace*{-0.4cm}
    \centering
    \hspace*{-10pt}\includegraphics[width=1.1\columnwidth]{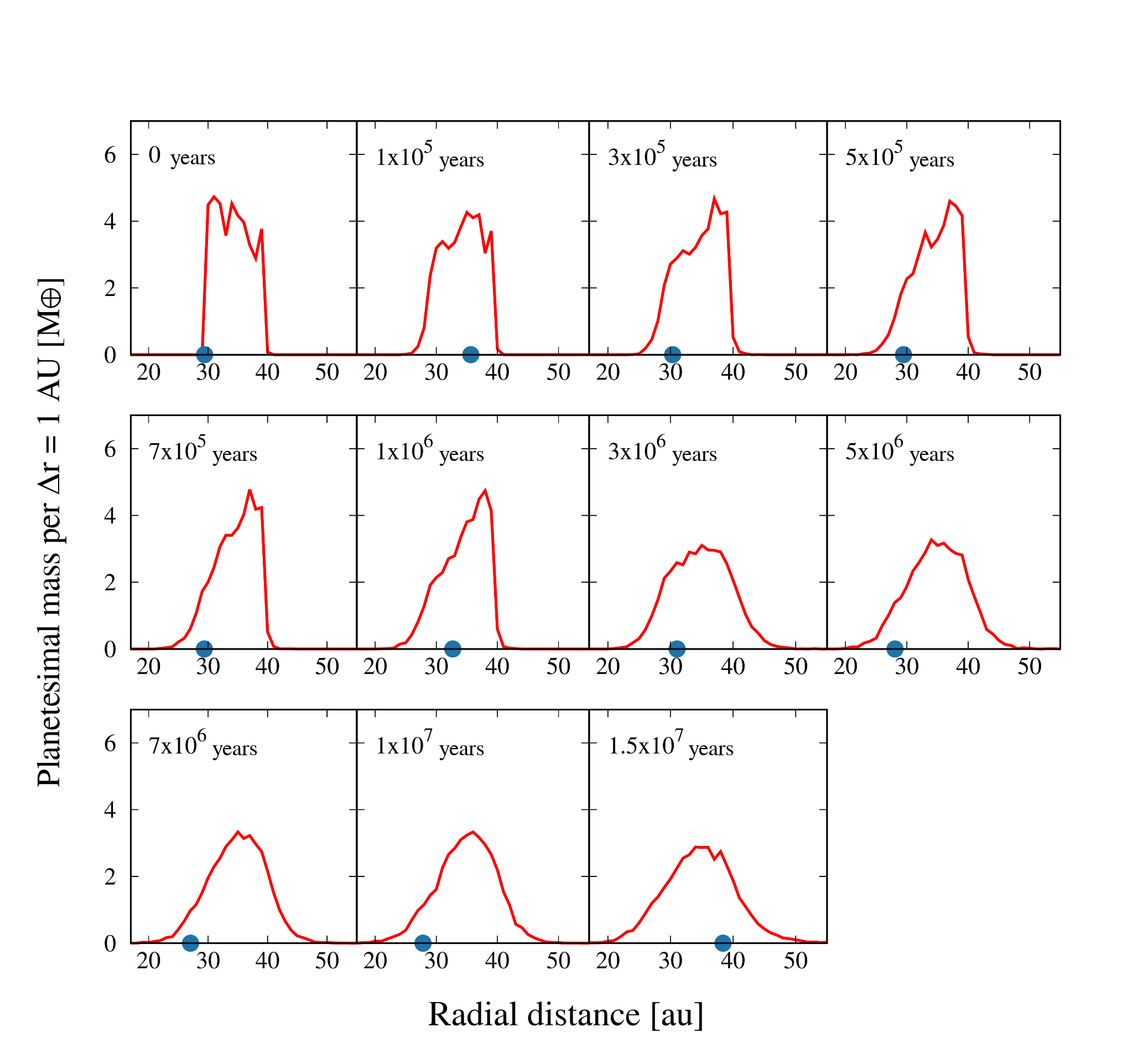}
    \vspace*{-0.7cm}
    \caption{Evolution of the radial mass distribution of planetesimals during the migration of a $1\,M_\oplus$ planet in a $40\,M_\oplus$ disk. The mass is computed in bins of width $1\,\mathrm{au}$. The blue dot indicates the distance of the planet from the star.}
    \label{fig:fig12}
\end{figure}

The maximum size of planetesimals $D_{\max}$ that are disrupted in collisions at the onset of the collisional cascade is shown in Fig.~\ref{fig:fig11}. This value reaches $40\,\mathrm{km}$ in the region between $30$ and $40\,\mathrm{au}$ (where the bulk of the disk mass is located; see Fig.~\ref{fig:fig12}).

\begin{figure}[h!]
 \vspace*{-0.4cm}
    \centering
    \hspace*{-10pt}\includegraphics[width=1.1\columnwidth]{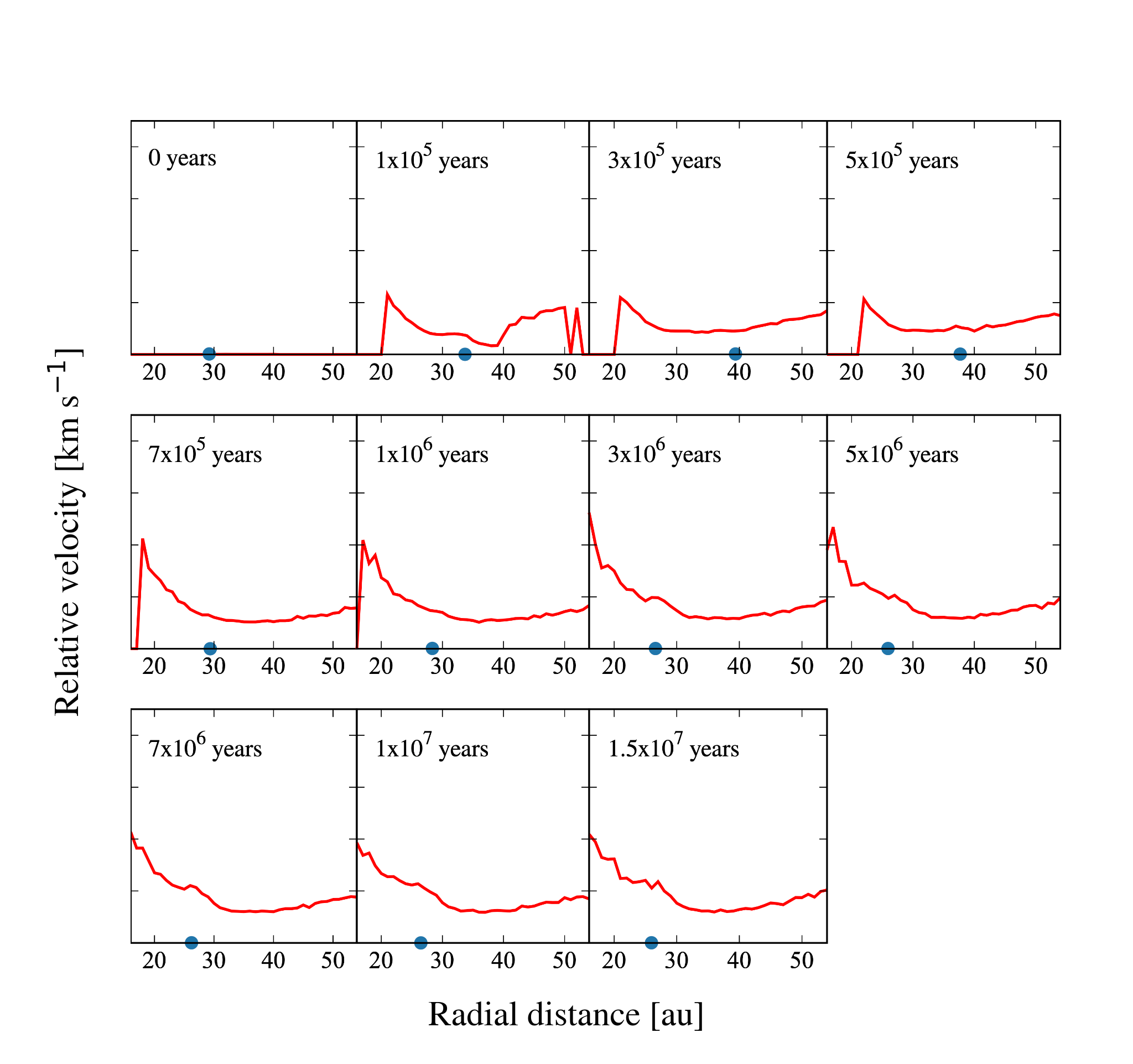}
    \vspace*{-0.7cm}
    \caption{Evolution of the relative velocity in the disk as a function of distance during the migration of a $5\,M_\oplus$ planet in a $40\,M_\oplus$ disk. The blue dot indicates the distance of the planet from the star.}
    \label{fig:fig13}
\end{figure}

We now consider the case of migration of a $5\,M_\oplus$ planet (yellow curve in the lower panel of Fig.~\ref{fig:fig3}). From $t = 7 \times 10^5$ yr onwards, the mass distribution in the disk (Fig.~\ref{fig:fig15}) and the relative velocities (Fig.~\ref{fig:fig13}) do not change significantly. We take this moment to mark the onset of the collisional cascade in the disk.

\begin{figure}[h!]
 \vspace*{-0.2cm}
    \centering
    \hspace*{-15pt}\includegraphics[width=1.0\columnwidth]{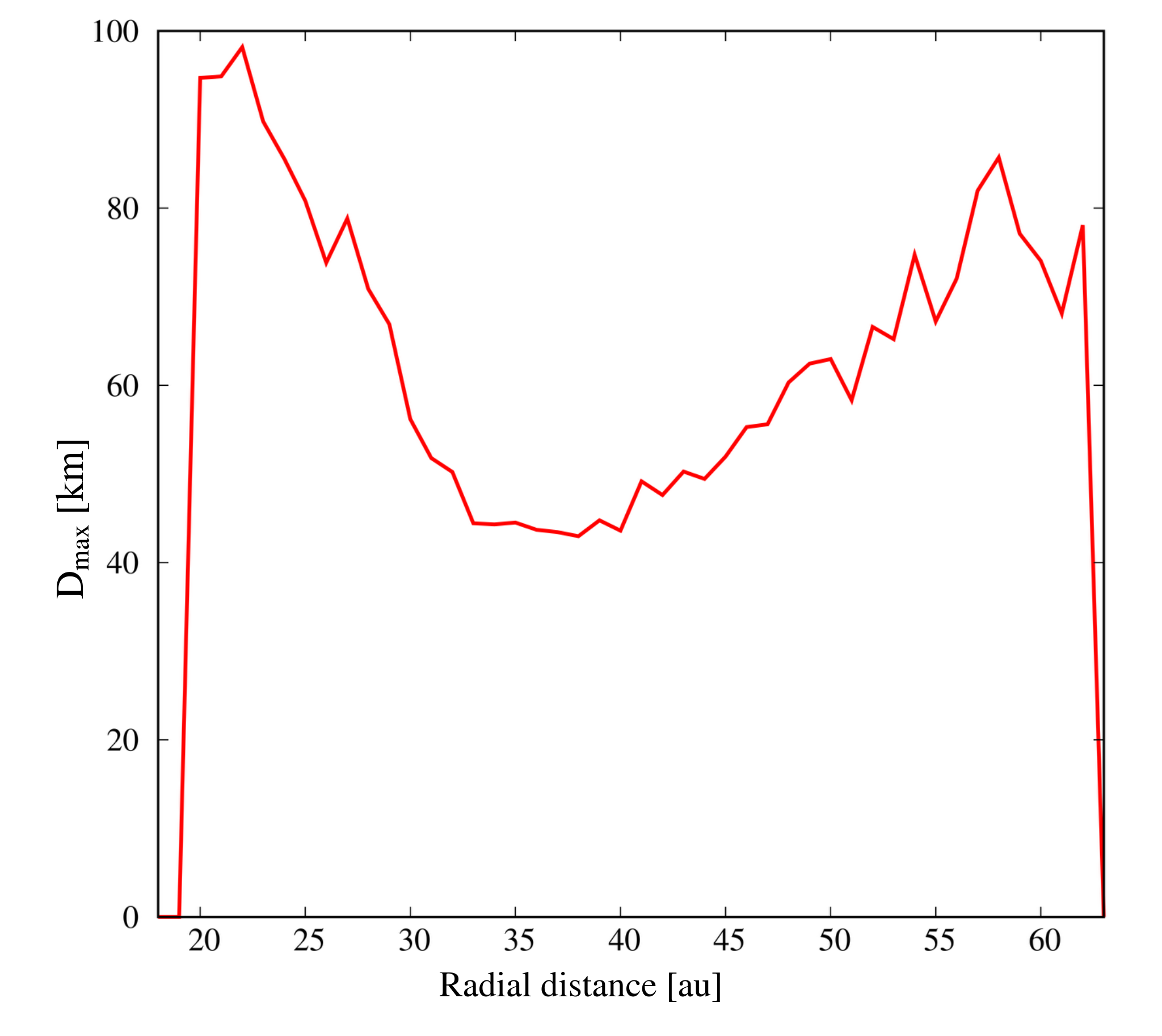}
    \vspace*{-0.2cm}
    \caption{Maximum planetesimal diameter $D_{\max}$ as a function of distance at $7 \times 10^5,\mathrm{yr}$ after the start of the integration for a $5\,M_\oplus$ planet migrating in a $40\,M_\oplus$ disk.}
    \label{fig:fig14}
\end{figure}

\begin{figure}[h!]
 \vspace*{-0.4cm}
    \centering
    \hspace*{-10pt}\includegraphics[width=1.1\columnwidth]{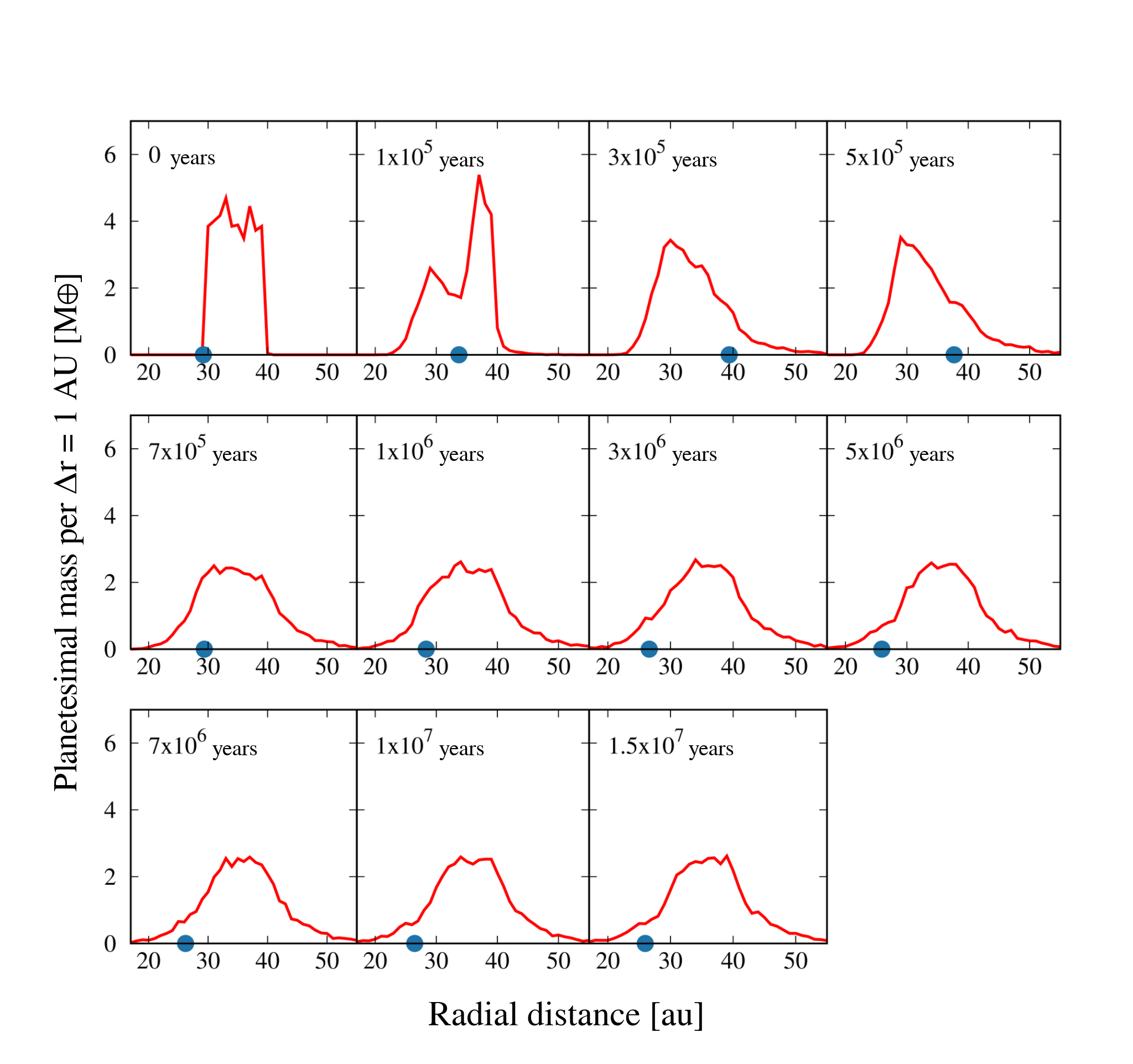}
    \vspace*{-0.7cm}
    \caption{Evolution of the radial mass distribution of planetesimals in the disk during the migration of a $5\,M_\oplus$ planet in a $40\,M_\oplus$ disk. The mass is computed in bins of width $1\,\mathrm{au}$. The blue dot indicates the distance of the planet from the star.}
    \label{fig:fig15}
\end{figure}

The maximum size of planetesimals $D_{\max}$ disrupted in collisions at this time ($t = 7 \times 10^5$ yr) is shown in Fig.~\ref{fig:fig14}. For the region of the disk containing the bulk of the mass (Fig.~\ref{fig:fig15}; the initial disk boundaries are $30$ and $40\,\mathrm{au}$), this value of $D_{\max}$ reaches $50\,\mathrm{km}$. This value does not significantly exceed that obtained for the case of migration of a $1\,M_\oplus$ planet.

\section{Discussion}

The interaction of Earth-mass planets with outer planetesimal disks differs significantly from the dynamical processes studied in detail for giant planets, in particular within the framework of the Nice model for the Solar System (see, e.g., \cite{Levison2007}). First, the migration of Earth-mass planets can occur in lower-mass disks and proceeds more rapidly than in the case of giant planets. Second, the degree of stochasticity in the motion of such planets is higher. As a result of these factors, and owing to the relatively low mass of the planet, planetesimals are only weakly captured into resonances \cite{Malhotra1995,Hahn1999,MurrayClay2006,Emelyanenko2010}, in contrast to the case of giant planets, and remain in the original disk while experiencing stronger perturbations during close encounters with the planet. Migration of Earth-mass planets can also occur over larger radial distances.

Our numerical experiments confirm the result of \cite{Kirsh2009}, namely that there is a tendency for the planet to migrate towards the star. However, since in our model the planet initially starts near the inner boundary of the planetesimal disk (in contrast to \cite{Kirsh2009}), it first moves away from the star. At a certain stage, the motion of the planet reverses and migration proceeds towards the star. The change in the direction of migration occurs at random times, determined by the chaotic redistribution of angular momenta of planetesimals whose orbits cross that of the planet.

\begin{figure}[h!]
 \vspace*{-0.4cm}
    \centering
    \hspace*{-10pt}\includegraphics[width=1.1\columnwidth]{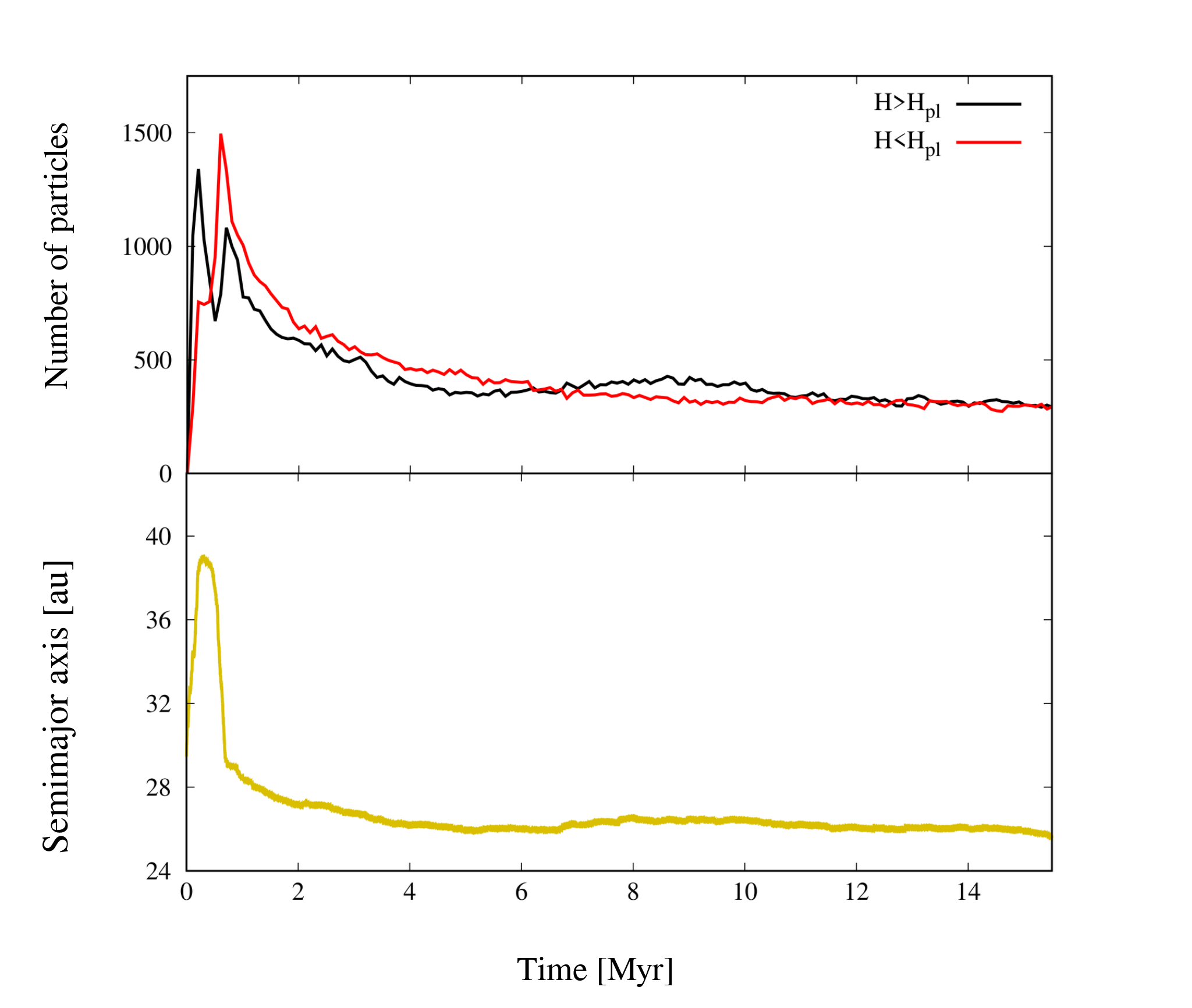}
    \vspace*{-0.7cm}
    \caption{Number of planetesimals with $H > H_p$ and $H < H_p$ (top panel) during the migration of a $5,M_\oplus$ planet in a $40,M_\oplus$ disk (bottom panel).}
    \label{fig:fig16}
\end{figure}

The direction of migration is determined by the relative number of planetesimals crossing the planet’s orbit with different values of the quantity
\begin{equation}
H = \sqrt{a \left(1 - e^2\right)} \cos i.
\end{equation}
If the number of planetesimals with $H > H_p$ (where $H_p$ is the value of $H$ for the planet) exceeds the number of planetesimals with $H < H_p$, the planet migrates deeper into the disk (i.e. away from the star), and vice versa \cite{Emelyanenko2007}. Figure~\ref{fig:fig16} shows how the number of such planetesimals (upper panel) correlates with the direction of migration at each moment of time (lower panel).

In some cases, the planet migrates through almost the entire disk, and the reversal of migration occurs near the outer boundary of the disk. We have also identified cases with a repeated change in the direction of migration near the inner boundary of the disk. As a rule, at the end of the migration process the planet remains on an orbit located a few au inward from the initial inner boundary of the disk, closer to the star.

The dynamical properties of the migration of an Earth-mass planet, in particular reversible migration, lead to the formation of a specific structure in the planetesimal disk. If the planet penetrates deeply into the planetesimal disk, its boundaries expand. As a result of the action of the migrating planet, the orbits of planetesimals are significantly perturbed (for a planet with a mass of $1\,M_\oplus$, the maximum eccentricities exceed $0.3$, and the inclinations exceed $10^\circ$). By the end of the migration process, the planetesimal disk is in an excited state (with a mean eccentricity of $\sim 0.1$ for a $1\,M_\oplus$ planet), while the planet is located far from the bulk of the disk.

As a result, a planetesimal disk is formed in which the relative velocities are sufficient to disrupt large planetesimals. According to our estimates, the maximum diameter of monolithic basaltic planetesimals for which a collisional cascade is initiated during the passage of an Earth-mass planet through the disk is $\sim 40\,\mathrm{km}$. In reality, this value depends on the material properties of planetesimals. For example, in the case of icy bodies, this size is several times larger for the same mean relative velocities \cite{Benz1999}.

There are also uncertainties in the initial masses of planetesimal disks. In our numerical experiments, we considered disk masses of $20$ and $40\,M_\oplus$, guided by estimates of the initial trans-Neptunian disk mass in the Solar System \cite{Nesvorny2012,Griveaud2024,Crida2009}. Naturally, in exoplanetary systems this mass may differ. Simple qualitative estimates \cite{Gomes2004,Levison2007,Ida2000} show that variations in this parameter primarily affect the mass of a planet capable of migrating within a planetesimal disk.

\section{Conclusion}

In this work, it is shown that a planet with a mass of the order of an Earth mass ($0.5\,M_\oplus$, $1\,M_\oplus$, and $5\,M_\oplus$), initially located near the inner boundary of a planetesimal disk, migrates into the disk. The depth of penetration of the planet into the disk is a random quantity determined by the distribution of angular momenta of planetesimals encountering the planet. However, at a certain stage, the direction of migration changes, and the planet returns to the inner boundary of the disk.

During such reversible migration, the planet perturbs the orbits of planetesimals and increases their relative velocities in the region of the disk traversed along its migration path. The relative velocities of planetesimals increase to values sufficient to disrupt large planetesimals in collisions. As a result, a collisional cascade is initiated, leading to the production of dust. It is shown that, during the passage of an Earth-mass planet through the outer planetesimal disk ($30$--$40\,\mathrm{au}$ in our model), the mean relative velocities in the bulk of the disk increase to values sufficient to disrupt monolithic basaltic planetesimals with sizes of $\sim 40\,\mathrm{km}$. In the case of similar migration of a $5\,M_\oplus$ planet, monolithic basaltic planetesimals with diameters up to $50\,\mathrm{km}$ can be disrupted.

Thus, the interaction of even a relatively low-mass planet (of order an Earth mass) with a planetesimal disk can lead to the production of dust particles observed in outer debris disks.

\section*{Acknowledgements}

The authors are grateful to the referee for helpful comments.

\bibliographystyle{unsrt}
\bibliography{bibfile}

\end{document}